\documentclass{cernyrep}
\usepackage{texnames}
\usepackage[T1]{fontenc}
\usepackage{varwidth}
\usepackage{xcolor}

\usepackage{graphicx}
\usepackage{amssymb, latexsym, mathrsfs}
\usepackage{amsmath}
\usepackage{color}
\usepackage{transparent}
\usepackage{multicol}
\usepackage{booktabs}
\usepackage[caption=false,font=footnotesize]{subfig}
\usepackage{multirow}

\newcommand{\figurefolder}{./figures}

\begin{document}

\title{Radiation Hardening of LED Luminaires for Accelerator Tunnels}

\author{J. D. Devine, and A. Floriduz}

\institute{CERN, Geneva, Switzerland}

\maketitle 

\begin{abstract}
This paper summarises progress made towards the radiation hardening of LED emergency luminaires for evacuation and emergency response within the underground areas of the CERN accelerator complex. The objective has been to radiation harden existing Commercial Off-The-Shelf (COTS) emergency luminaires to maximise lighting performance, without compromising IEC 60598-2-22 compliance. A systems level approach has been adopted, leading to the development of a diode bridge based AC/DC power converter. Modified COTS luminaires including this converter design have been irradiated (to 100~kGy TID using a Cobalt-60 source), with results of a subsequent photometric analysis presented. Following encouraging tests results, a reference design for the power converter has been released under the CERN Open Hardware License to encourage manufacturer adoption. The paper concludes with areas of interest for future research in further improving the radiation hardness of LED emergency lighting for accelerators with detailed studies at the component level for high power white LED devices and associated optical components.\\\\
{\bfseries Keywords}\\
Radiation resistant power supply; emergency lighting; particle accelerators.
\end{abstract}

\section{Introduction}
CERN has around 45~km of underground tunnels containing particle accelerators, beam transfer lines and access tunnels at depths of up to 100~m below the surface \cite{saraiva2013radiation}. During accelerator shutdown periods extensive access is required for human operatives performing maintenance and upgrade activities. A robust emergency lighting system, comprising in excess of 3000 luminaires, exists to ensure safe evacuation from underground areas in the event of an emergency. Currently emergency lighting is provided by legacy lighting systems dating from the original underground accelerator installations (Super Proton Synchrotron and Large Electron Positron Collider) which comprise a mixture of low-pressure sodium and incandescent luminaires. These are now increasingly difficult to maintain due to regulations applicable to lighting systems which limit the supply of replacement lamps, creating the need for the development of a replacement system based on current lighting technology and radiation resistant properties.

\subsection{Emergency lighting systems}
Emergency lighting systems in tunnel accelerators are intended to provide a minimum level of illumination for escape routes, in case of a loss of power to normal luminaires. The purpose of emergency lights is to allow people working in underground areas to escape safely from the tunnels and to facilitate the intervention of emergency services if required. 
At CERN, emergency luminaires are supplied from rechargeable Nickle Metal Hydroxide (NiMh) batteries via 48~V DC to 230~V AC inverters. Luminaires are supplied through fire rated cables, with each underground area typically being served by two separate electrical circuits to provide redundancy. In most areas of the Large Hadron Collider (LHC) tunnel, the emergency lighting system makes use of low-pressure sodium discharge lamps retained from the era of the LEP accelerator. Sodium discharge lamps with wire wound inductive ballasts have proved to be highly reliable and radiation resistant throughout the radiation environments created by both the LEP and the LHC \cite{roed2012method}, as they are intrinsically highly tolerant to radiation effects.

In recent years, Light-Emitting Diodes (LED) technology has become an attractive alternative to traditional incandescent, fluorescent and discharge lamps, not only for the general illumination market but also in the field of emergency lighting, thanks to their compact dimensions, instant response and high luminous efficacy. A major role in this transition has been played by national and international regulations, such as the EU directive no. 245/2009, which has led to a ban on certain types of inefficient metal halide, mercury and  high-pressure sodium lamps. 
As a combined consequence of legislative pressure and technological progress, the expansion of the LED market appears inevitable while high intensity discharge lamps, including low-pressure sodium lights, are losing market share and ultimately becoming obsolete \cite{de2014solid}.
For this reason, a new generation of LED emergency luminaires is being developed for use in CERN underground facilities.

\subsection{Radiation environment}
\label{subsection:radEnvironment}
The LHC radiation environment is characterised in \cite{roed2012method}, \cite{de2015radiation} and \cite{r2e2016}. 
Since the emergency luminaires will eventually be installed on the walls of  all underground areas (tunnels and caverns) of the CERN accelerator complex, they must be radiation hardened to an appropriate level in order to prevent premature failure. 
Working within the context of the As Low As Reasonably Achievable (ALARA) principle, the system must be designed to maximise operational lifetime, in order to prevent interventions related to luminaire breakdown and replacement. For this reason, the desired lifetime of emergency lighting luminaires is a minimum of 5 years. 

The latest values (2015) of yearly  High Energy Hadrons (HEH) fluence (i.e. fluence of all hadrons with an energy  greater than 20~MeV), 1~MeV equivalent neutron (neq) fluence and  Total Integrated Dose (TID) in the LHC tunnel and its injectors are presented in \cite{de2015radiation}; in addition, this report illustrates that TID values on the tunnel walls are approximately one order of magnitude smaller than close to the corresponding beam lines, as evidenced by from both direct TID measurements and Monte Carlo simulations. Therefore, a suitable target dose for the emergency luminaires under development can be set to 100~kGy, since this value exceeds the expected TID relative to 5 years irradiation in most of the CERN accelerator complex, meeting the requirement of the desired lifetime.

In underground areas, an LED emergency lighting system may also be interlocked to activate only during machine stops (i.e. no beam condition), since no access to these areas is possible for human operators during accelerator runs.%
\footnote{Due to the long warm-up time of low-pressure sodium lamps, the present emergency luminaires in the CERN accelerator complex are always energised in order to avoid conflict with EN 1838 requirements for the rapid activation of emergency lighting systems. Therefore, adoption of LED lighting  provides increased flexibility for energy savings thanks to potential for reduced operational time.} 
Hence, LED emergency luminaires in CERN tunnels and caverns will also be studied when not powered, at ambient temperature (for reference, typical LHC tunnel temperature during physics runs is within the range  $17\pm3^{\circ}$C).

\begin{figure}
	\centering
	\includegraphics[scale=0.285]{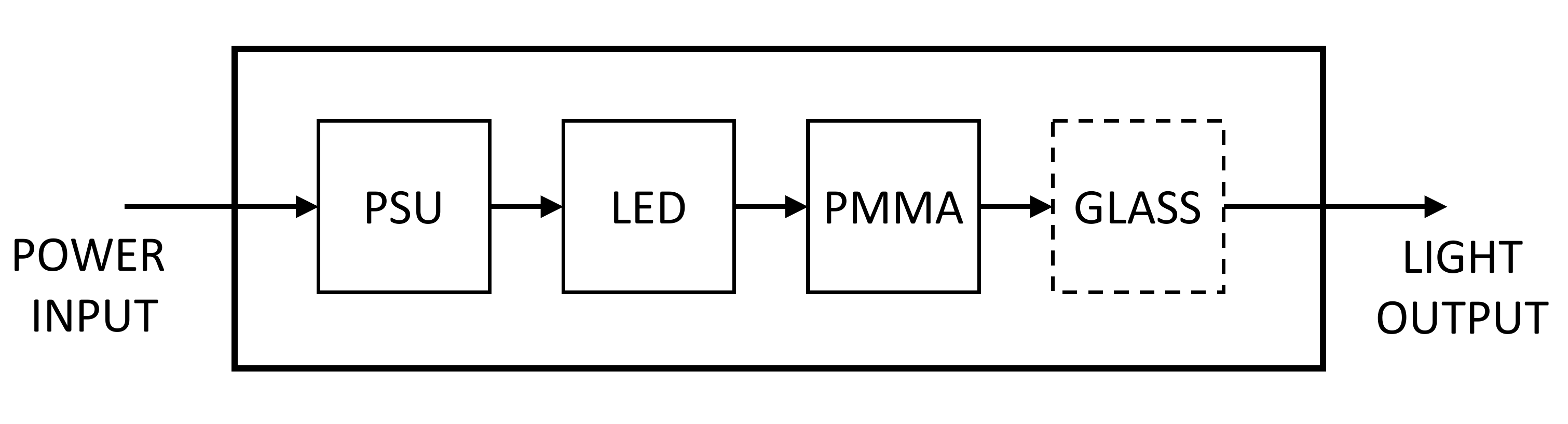}
	\caption{System model of an LED emergency luminaire.}
	\label{fig:model}
\end{figure}

\subsection{LED luminaires}
The block diagram model of a COTS LED emergency luminaire is depicted in Fig. \ref{fig:model}. The system comprises a number of components, including a Power Supply Unit (PSU), which converts the input supply voltage (typically 230~V 50~Hz AC) to the DC value required by the LEDs supplied. Switch-Mode Power Supply (SMPS) topologies, such as buck, boost and flyback converters, are commonly used to supply LEDs \cite{rico2005evaluation} in COTS luminaires. However, these PSUs are not designed to be radiation hard and when tested show very poor performance, causing almost immediate catastrophic failure of LED luminaires when subjected to a typical accelerator radiation environment \cite{adell2002total}. The reliability required for their use in accelerator environments (see, for example, \cite{adell2013radiation} for a review of radiation effects on power converters), can therefore not be guaranteed. Each luminaire may contain one or more high power LED light sources, each with a lumen output of approximately 100~lm or greater. The effect of radiation on LED light output has been studied extensively for optical power levels up to 100~mW \cite{johnston2013radiation}, \cite{jimenez2006proton}, \cite{pearton2016review} however little published research exists for high power LEDs up to 1~W and beyond. The individual LED is generally equipped with a plastic lens to improve the light output distribution (see Fig. \ref{fig:model}); the lens is usually made of PolyMethyl MethAcrylate (PMMA) plastic selected for high light transmission, a property which degrades when exposed to radiation \cite{toh2011pmma}. Depending upon the luminaire design, the LED may also be contained in an IP6X housing protected with glass windows to allow light transmission (see Fig. \ref{fig:model}). Borosilicate retains its optical transmission properties relatively well when irradiated \cite{baydogan2012borosilicate} and has been specified for installations following the tests described in this paper. However, the overall light output of the system may be further improved by careful selection of quartz or high purity fused silica materials \cite{esamats} in the future.

Having developed a simple system level model for the LED luminaires, the failure modes of each component and the corresponding impact on the system performance can be demonstrated (see Table \ref{table:r2e}). Early functional testing on a range of COTS products rapidly indicated that first stage of radiation hardening would be to prevent the catastrophic failure mode of the PSU due to the inclusion of sensitive SMPS. Following this insight, CERN has worked with two luminaire vendors to modify their standard products into radiation hardened versions. The luminaires contain modified PSUs, based on a diode bridge rectifier principle, providing significantly increased radiation hardness, and do not conflict with the requirements of IEC 60598-2-22.

\begin{table}[!h]
	\renewcommand{\arraystretch}{1.3}
	\caption{Radiation effects on luminaire components.}
	\label{table:r2e}
	\centering
	\begin{tabular}{ll}
		\toprule
		Component & Effects of radiation\\ \midrule
		PSU & Catastrophic failure modes for SMPS \\ \midrule
		LED & \begin{tabular}{@{}l@{}} 
			Radiation damage (displacement) leading to reduced \\ flux \end{tabular} \\ \midrule
		\begin{tabular}{@{}l@{}} PMMA (plastic \\ lens compound)  \end{tabular}  & \begin{tabular}{@{}l@{}} Free radical formation leading to reduced light \\ transmission \\ 
			Degraded mechanical properties \end{tabular} \\ \midrule
		Glass & \begin{tabular}{@{}l@{}} Colour centre formation leading to reduced light \\ transmission\end{tabular} \\ \bottomrule
	\end{tabular}
\end{table}

\section{Design response}

\begin{figure}
	\centering
	\subfloat[]{\includegraphics[width=0.88\columnwidth]{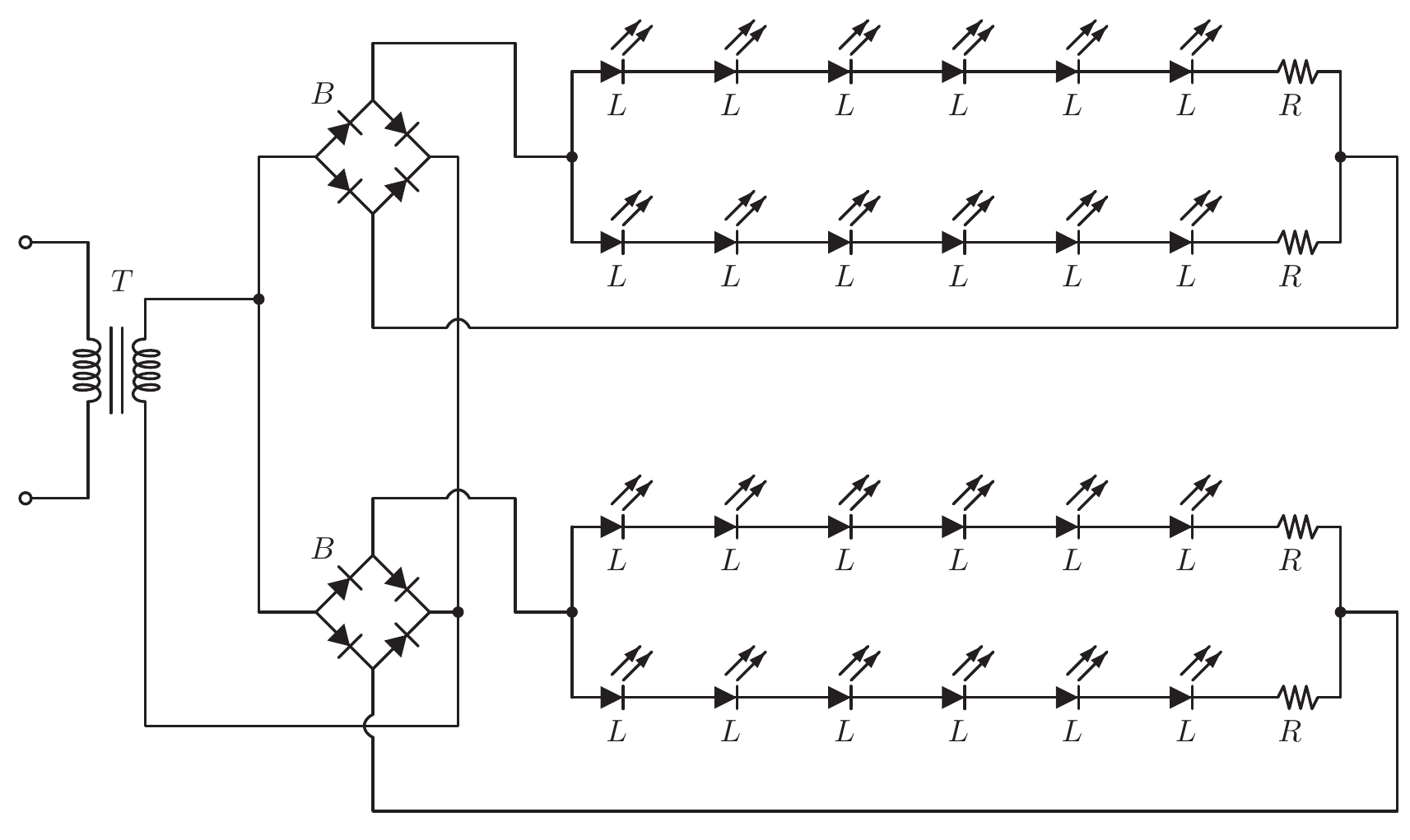}\label{fig:COTSdiode}}
	\vspace{1.2cm}
	\subfloat[]{\includegraphics[width=0.97\columnwidth]{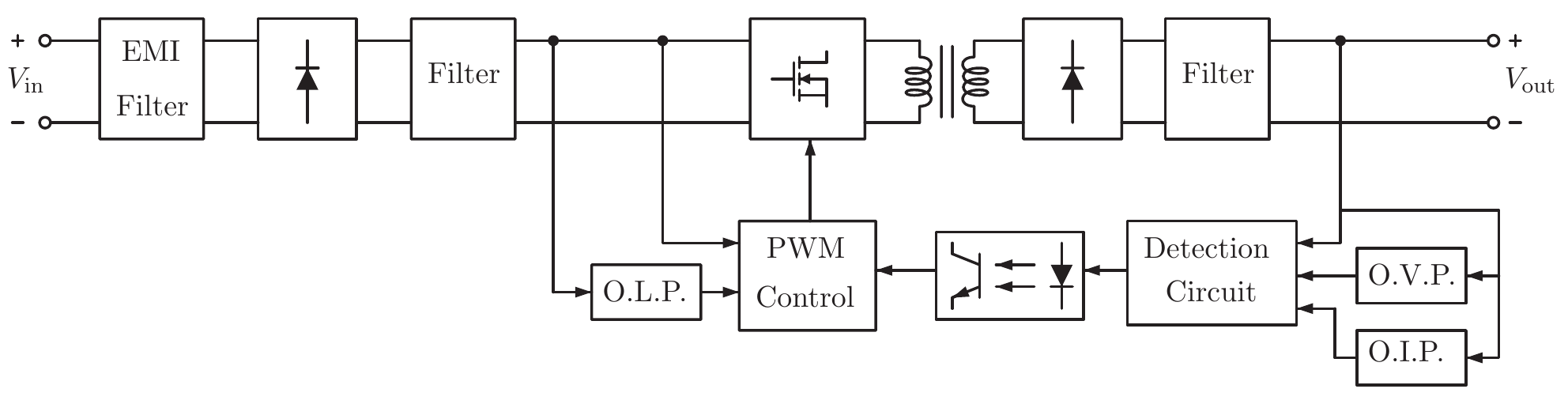}\label{fig:SMPS}}
	\caption{(a) Circuit diagram of the power supply using diode bridge rectifier topology. (b) Block diagram of the SMPS used in COTS luminaires under test.} 
\end{figure}

\subsection{Preliminary irradiation tests}
\label{sec:preliminary}
Preliminary irradiation tests of two sets of COTS emergency lighting systems have been performed at CNRAD facility \cite{CNRAD2016} at CERN. The goal was to analyse the response to radiation of COTS power supplies in order to identify a suitable basis for future design of a rad-tolerant PSU for LED lighting.  The first set of luminaires under test comprised three lamps with a simple power supply based on the principle of a transformer and a diode bridge rectifier, whose circuit diagram is represented in  Fig. \ref{fig:COTSdiode}. Each luminaire presents a step down transformer (denoted with $T$ on the circuit diagram) supplying two W08M diode bridge rectifiers (manufactured by MIC Electronics, and denoted with $B$), each one feeding two parallel strings of six series connected low power, low current ($<$50~mA) white LEDs, indicated as $L$. Each string is series connected to a 560~$\Omega$ resistance in order to prevent thermal runaway of LEDs. The second batch comprised three luminaires using SMPS converters, whose block diagram is depicted in Fig. \ref{fig:SMPS}. As can be seen, the AC input voltage is first rectified and filtered; an EMI filter is used to mitigate the injection of harmonics into the grid. Then, this stage is followed by a high frequency DC/DC converter composed of an inverter using MOSFET power switches, an high frequency transformer and an AC/DC diode rectifier followed by a filter, to smooth the DC voltage. The power switches of the inverter are controlled via PWM modulation, at a switching frequency of 60~kHz. The control feedback loop is closed by the measurement/detection system, which sends  backs to the control circuitry the DC output voltage, using opto-couplers. In order to prevent overloads on the input and output side, an overcurrent protection system (respectively denoted as O.L.P. and O.I.P. on the input and output side) is embedded into the control system; moreover, and Over Voltage Protection (O.V.P.) on the DC output side is also present.

\begin{figure}
	\centering
	\subfloat[]{\includegraphics[height=7cm]{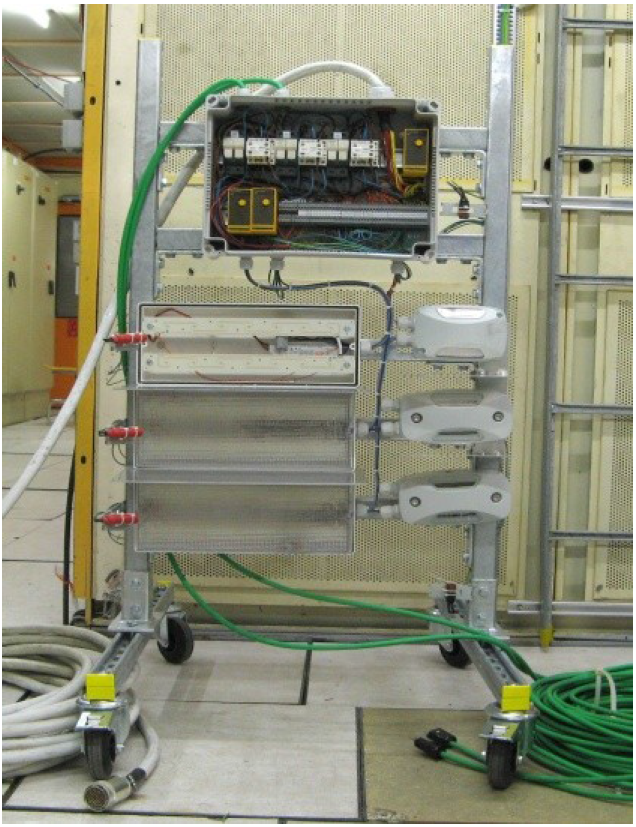}%
		\label{fig:preliminary_layout}}
	\hfil
	\subfloat[]{\includegraphics[height=7cm]{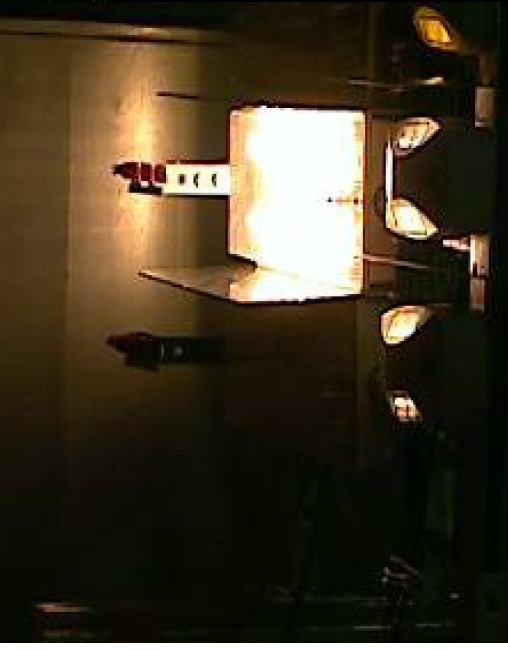}%
		\label{fig:preliminary_irrad}}
	\caption{(a) Layout of luminaires during preliminary irradiation tests at CNRAD (CERN); on the left, three luminaires using SMPS, on the right three luminaires with diode bridge rectifiers. (b) Picture from remote camera after 5 hours of irradiation; the three luminaires using diode bridges are on the right, luminaires with SMPS on the left.}
	\label{fig:preliminary}
\end{figure}

The irradiation took place into four steps, with cumulative TID of 9, 47.9, 93.7 and 129.3~Gy. The four irradiation steps lasted respectively 230, 900, 1260 and 1095~hours,  separated by intervals with no irradiation of 109, 75 and 105~hours. The irradiation conditions are summarized in Table \ref{table:COTS_PSU}.
	The six luminaires were installed on a test bench (illustrated in Fig. \ref{fig:preliminary_layout}), located in CNRAD test area, with a deported RadMon \cite{masi2012radmon} positioned in the middle of the rack, at about 60~cm from ground, to measure precisely the radiation level of the devices under test. All luminaires were powered while irradiated (active test) and were monitored through a remote rad-hard camera.

\begin{table}[!h]
	\centering
	\caption{Irradiation condition of the COTS luminaires.}
	\label{table:COTS_PSU}
	\renewcommand{\arraystretch}{1.3}
	\begin{tabular}{lccccc}
		\toprule
		& Step 1 & Step 2 & Step 3 & Step 4 & Total \\ \midrule
		TID (Gy) & 9 & 38.9 & 45.8 & 35.6 & 129.3\\
		1 Mev neq fluence $\left(10^{11}\cdot\mbox{cm}^{-2}\right)$ & 0.7 & 3.0 & 3.6 & 2.7 & 10.0  \\
		HEH fluence $\left(10^{11}\cdot\mbox{cm}^{-2}\right)$ & 0.5 & 2.1 & 2.5 & 1.9 & 7.0 \\
		Duration (h)  & 230 & 900 & 1260 & 1095 & 3485 \\
		Time after previous step (h)  & / & 109 & 75 & 105 & /\\ \bottomrule  
	\end{tabular}
\end{table}

The three luminaires using SMPS units failed just after the start of the test, after 1 hour, 4 hours and 18~hours of irradiation, corresponding to TID values of less than 0.1~Gy, 0.1~Gy and 0.4~Gy respectively. These results are summarized in Table \ref{table:SMPS_irrad}. 
	Figure \ref{fig:preliminary_irrad} shows the test bench from the remote camera after 5 hours of irradiation, with two of the luminaires using SMPS already failed. The failure of SMPS luminaires can be ascribed to displacement damage effects on the power switches of the high frequency DC/DC converter, on the PWM control circuitry and on opto-couplers.

\begin{table}[!h]
	\centering
	\renewcommand{\arraystretch}{1.5}
	\caption{Irradiation of luminaires with SMPS.}
	\label{table:SMPS_irrad}
	\begin{tabular}{lccc}
		\toprule
		Characteristics & Luminaire 1 & Luminaire 2 & Luminaire 3 \\ \midrule
		\begin{tabular}{@{}l@{}} Time before failure (h) \end{tabular}  & 1 & 4 & 18 \\ 
		\begin{tabular}{@{}l@{}} TID reached (Gy) \end{tabular}  & $<0.1$ & 0.1 & 0.4 \\
		\begin{tabular}{@{}l@{}} HEH fluence $\left(\mbox{cm}^{-2}\right)$\end{tabular}  & $<5.4\cdot 10^8$ & $5.4\cdot 10^8$ & $2.2\cdot10^9$ \\
		\begin{tabular}{@{}l@{}} 1 MeV neq fluence $\left(\mbox{cm}^{-2}\right)$ \end{tabular}  & $<7.3\cdot10^8$ & $7.3\cdot10^8$ & $2.9\cdot10^9$ \\ \bottomrule
	\end{tabular}
\end{table}

Luminaires using power supplies with diode bridge rectifiers succeeded in completing the irradiation test, working for the full duration without showing any significant degradation in illuminance, as confirmed by visual inspections and luxmeter measurements performed after the end of the irradiation. The total dose reached was of 129~Gy, at a HEH fluence of $7.0\cdot10^{11}\,\mbox{cm}^{-2}$ and 1 MeV neq fluence of $1.0\cdot10^{12}\,\mbox{cm}^{-2}$.

\begin{figure}[!h]
	\centering
	\subfloat[]{\includegraphics[width=0.48\columnwidth]{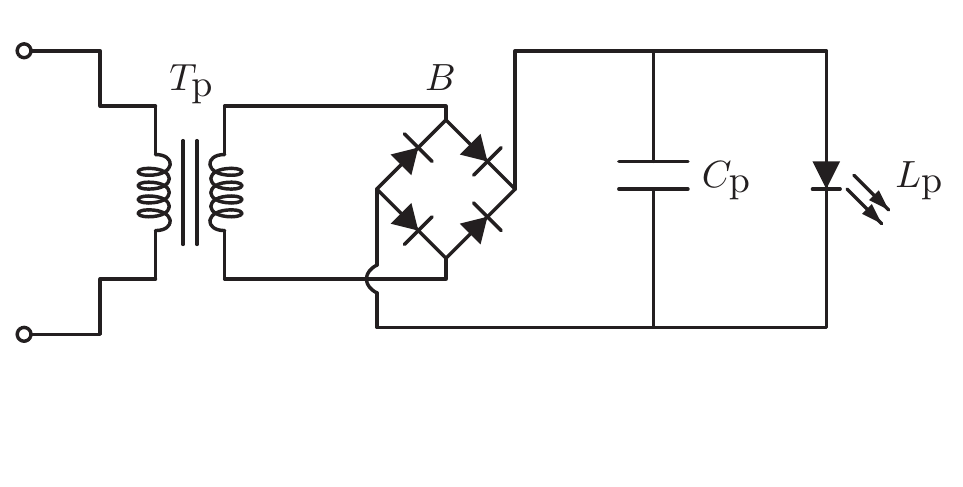}%
		\label{fig:proto_diagram}}
	\hfil
	\subfloat[]{\includegraphics[width=0.4\columnwidth]{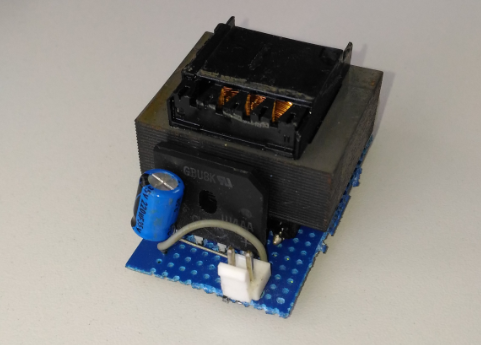}%
		\label{fig:proto_pic}}
	\caption{(a) Circuit diagram of the prototype power supply using GBU8K diode bridge rectifier. (b) Picture of the prototype PSU assembled on a stripboard.}
	\label{fig:proto}
\end{figure}

	The irradiation test identified the power supply as the  most critical component in COTS emergency luminaires and confirmed that converters using diode bridge topology offer an higher reliability in radiation environment than SMPS. Following these preliminary results, a prototype of a rad-tolerant PSU using the bridge rectifier topology has been built; this converter includes a GBU8K diode bridge, samples of which have been previously tested for radiation hardness by the LHCb experiment \cite{bager2002lhcb}, up to 1 MeV neq fluence of $4.8\cdot 10^{13}\,\mbox{cm}^{-2}$, corresponding to a TID of 3.3~kGy; GBU8K bridge rectifier makes use of glass passivated junctions.
	The circuit diagram of the prototype power supply is shown in Fig. \ref{fig:proto_diagram}; it includes a 230:6 step-down transformer (denoted with $T_{\scriptsize{\mbox{p}}}$ in Fig. \ref{fig:proto_diagram}) and an electrolytic 220~$\mu$F capacitor (denoted with $C_{\scriptsize{\mbox{p}}}$). The power supply feeds a single Cree XR-E LED, indicated with $L_{\scriptsize{\mbox{p}}}$, producing white light; Cree XR-E LED is in turn composed of a blue InGaN LED and yellow phosphors. A picture of an assembled prototype is shown in Fig. \ref{fig:proto_pic}. 2 PSU prototypes have been tested in CNRAD facility, in a test position which allowed higher TID and fluences than the location of the previous test of COTS luminaires.  Irradiation took place in four steps with cumulated TID of 160.2, 406.1, 760.1 and 1107.6~Gy, with   final cumulated HEH fluence of $1.1\cdot10^{13}\,\mbox{cm}^{-2}$ and cumulated 1 MeV neq fluence of $7.7\cdot10^{12}\,\mbox{cm}^{-2}$. From a comparison with the values presented in \cite{de2015radiation}, the TID reached during this irradiation test corresponds approximately to the dose accumulated on the LHC tunnel walls after 5 years exposure.
	All the PSUs were powered while irradiated (active test).
	Inspections were performed after the last irradiation step, and confirmed that all PSUs were still working. Following this encouraging result, the prototype has been chosen as the basis for a rad-tolerant PSU for LED lighting.

\begin{table}
	\centering
	\caption{Irradiation condition of the prototype PSU.}
	\label{table:protoPSU}
	\renewcommand{\arraystretch}{1.3}
	\begin{tabular}{lccccc}
		\toprule
		 & Step 1 & Step 2 & Step 3 & Step 4 & Total \\ \midrule
		TID (Gy) & 160.2 & 245.9 & 354 & 347.5 & 1107.6\\
		1 Mev neq fluence $\left(10^{12}\cdot\mbox{cm}^{-2}\right)$ & 1.6 & 2.4 & 3.5 & 3.4 & 11  \\
		HEH fluence $\left(10^{12}\cdot\mbox{cm}^{-2}\right)$ & 1.1 & 1.7 & 2.5 & 2.4 & 7.7 \\ \bottomrule
	\end{tabular}
\end{table}

\subsection{Power Supply Modifications}
\begin{figure}[!h]
	\centering
	\subfloat[]{\includegraphics[width=0.57\columnwidth]{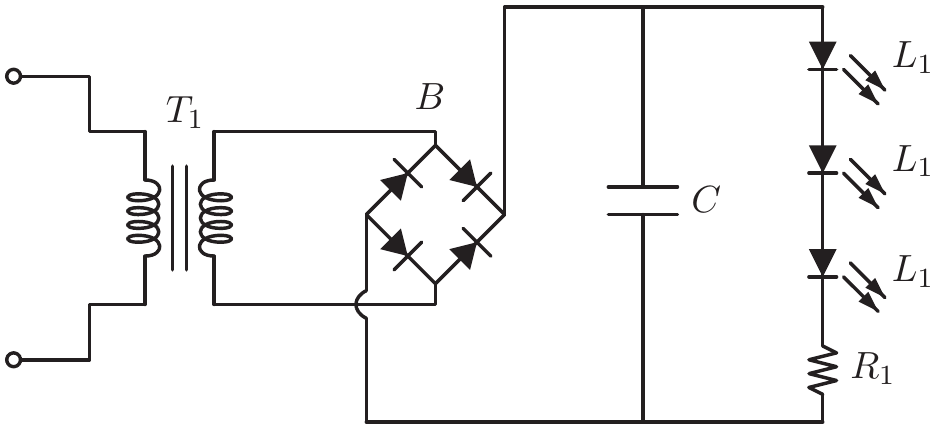}%
		\label{fig:circuitdiagram1}}
	\hfil
	\subfloat[]{\includegraphics[width=0.57\columnwidth]{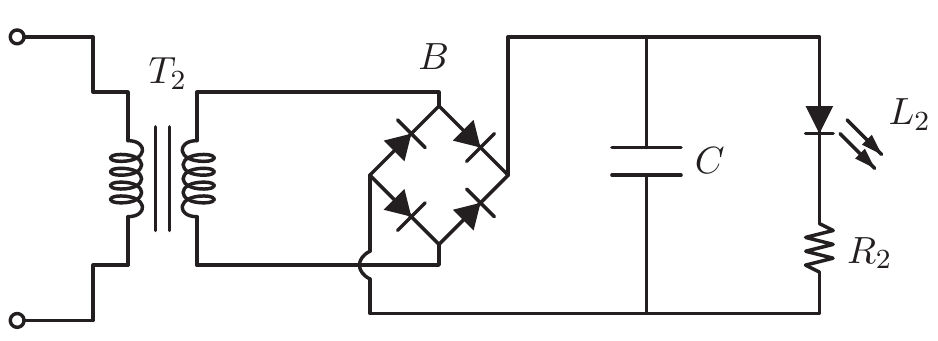}%
		\label{fig:circuitdiagram2}}
	\caption{(a) Circuit diagram of Vendor 1 power supply. (b) Circuit diagram of Vendor 2 power supply.}
	\label{fig:circuitdiagrams}
\end{figure}

Once a suitable power converter topology had been identified, work with luminaire vendors was commenced in order to incorporate the new power converter topology into standard products with optical characteristics suitable for emergency evacuation lighting within underground tunnels. A second vendor with an alternative luminaire design more suited to underground caverns and larger spaces was also approached. The power converter specification was issued to both vendors, which included the same bridge diode rectifier (GBU8K) used in the prototype PSU presented in Section \ref{sec:preliminary}. As a result of these initiatives, two types of radiation hardened emergency luminaire were produced for further radiation testing.
The circuit diagrams of Vendor 1 and Vendor 2 power supplies are shown respectively in Fig. \ref{fig:circuitdiagram1} and Fig. \ref{fig:circuitdiagram2}. The values of the components are listed in Table \ref{table:components}.

\subsection{Components under test}
The luminaires from Vendor 1 utilise three Cree XP-G LEDs as light sources, with a combination of PMMA lenses and glass windows. Vendor 2 uses a single Cree XR-E LED with a PMMA lens to direct the light output. Cree XR-E and XP-G LEDs produce white light; they are composed of blue LEDs (InGaN epitaxial layer on silicon substrate) with phosphors producing yellow light. To the best of our knowledge, no previous irradiation tests of Cree XR-E and XP-G LEDs have been reported. Both luminaires use the GBU8K diode manufactured by Vishay.  Further components include surface mounted multi-layer ceramic capacitors and metal-film resistors. 

\begin{table}[!h]
	\centering
	\renewcommand{\arraystretch}{1.3}
	\caption{Electrical components of Vendor 1 and Vendor 2 power supplies.}
	\label{table:components}
	\begin{tabular}{lllll}
		\toprule
		Name & Quantity & & Name & Quantity \\\midrule
		$T_1$ &  16~VA transformer, 230$\,:\,$14 &\qquad\qquad$\;\;$ & $C$ &  1~mF ceramic capacitor \\
		$T_2$ &  16~VA transformer, 230$\,:\,$7 & & 	$L_1$ &  Cree XR-E white LED \\
		$R_1$ &  16~$\Omega$ 5\% resistor, 25~W & & $L_2$ &  Cree XP-G white LED \\
		$R_2$ &  5~$\Omega$ 5\% resistor, 25~W  & & $B$ & GBU8K diode bridge rectifier \\\bottomrule 
	\end{tabular}
\end{table}

\section{Experimental results}

\subsection{Test methodology}
Gamma ray and neutron irradiation was performed on luminaires from both vendors. Gamma ray irradiation was conducted at the Fraunhofer Institute INS (Germany) to a TID of 100~kGy; two sample luminaires for each vendor were sent to Fraunhofer for test purposes. Neutron irradiation was performed at CEA in France on another set of samples, likewise composed of two luminaires from Vendor 1 and two luminaires from Vendor 2. Analysis of the neutron irradiated samples will be presented in a future work.
In addition to irradiation tests, a trial installation using the proposed LED luminaires has recently been completed in part of the SPS accelerator complex (TDC2 and TCC2 tunnels and target caverns) with a very challenging radiation environment \cite{Strabel2015}, with half of the luminaires permanently on and half interlocked to activate only when there is no beam present in the accelerator.

The gamma ray irradiation has been performed at the BGS low dose test facility at Fraunhofer, using a Cobalt-60 source at a dose rate of 720~Gy(Si)/h, to a final dose value of 100~kGy(Si). 
The luminaires were not powered during irradiation (passive test). As detailed in Section \ref{subsection:radEnvironment}, in CERN tunnels LED luminaires may be powered off while irradiated; for this reason, passive irradiation tests are also representative of actual radiation effects in accelerator tunnel environment. 
The irradiation was conducted in five steps, with final dose per exposure of 4.1, 15.3, 34.2, 50.0 and 100.0~kGy(Si). 
After each step (and within three hours after the completion of the exposure),  preliminary electrical and optical measurements were taken to verify the continued functioning of the luminaire.
For each step, this activity took less than two hours.
The irradiation of the sample luminaires lasted overall 146.7 hours. The time interval between the end of a step and the beginning of the following one was within 5 and 20 hours. The salient data of the gamma ray irradiation test are summarized in Table \ref{table:irradiation}. 
\begin{table}
	\renewcommand{\arraystretch}{1.3}
	\caption{Summary of gamma ray irradiation conditions.}
	\label{table:irradiation}
	\centering
	\begin{tabular}{ll}
		\toprule
		Parameter & Value\\ \midrule
		Irradiation facility & BGS low dose \\ 
		Irradiation source & Co-60 gamma \\ 
		Dose steps  & 4.1, 15.3, 34.2, 50.0, 100.0 kGy(Si) \\ 
		Dose rate &  720 Gy(Si)/h \\ 
		Time of irradiation & 146.7 h\\ 
		Time between steps & 9, 8, 5, 20 hours \\ \bottomrule
	\end{tabular}
\end{table}

\subsection{Results from goniophotometer analysis}
Following the gamma irradiation exposure, the luminaires were sent for a Type C Moving Mirror goniophotometer analysis at TSI-LUX Ltd. in the UK.
Electrical and photometric measurements have been performed and the results are collected in Table \ref{table:lux}. 

\begin{table}[!h]
	\renewcommand{\arraystretch}{1.3}
	\caption{Summary of electrical and photometric measurements.}
	\label{table:lux}
	\centering
	\begin{tabular}{lcc}
		\toprule
		Quantity & Vendor 1 & Vendor 2\\ \midrule
		Frequency & 50 Hz& 50 Hz \\ 
		Voltage & 230.06 V & 230.04 V \\ 
		Current & 0.031 A & 0.031 A\\
		Power & 6.50 W & 6.20 W \\
		Power factor & 0.924 & 0.879 \\
		Apparent power & 7.04 VA & 7.09 VA \\
		CRI (Ra) &  71 & 65 \\ 
		Luminous flux pre-irradiation & 900 lm & 209 lm \\ 
		Luminous flux post-irradiation & 177 lm & 55 lm \\ 
		\begin{tabular}{@{}l@{}} Variation in luminous flux \end{tabular}  & -80.33\% & -73.68\% \\ 
		Luminous efficacy pre-irradiation & 128.57 lm/W & 32.15 lm/W\\
		Luminous efficacy post-irradiation & 27.23 lm/W & 8.87 lm/W\\
		\begin{tabular}{@{}l@{}} Variation in luminous efficacy \end{tabular}  & -78.82\% & -72.41\%\\ \bottomrule
	\end{tabular}
\end{table}

\begin{figure}[!b]
	\centering
	\subfloat[]{\includegraphics[height=5cm]{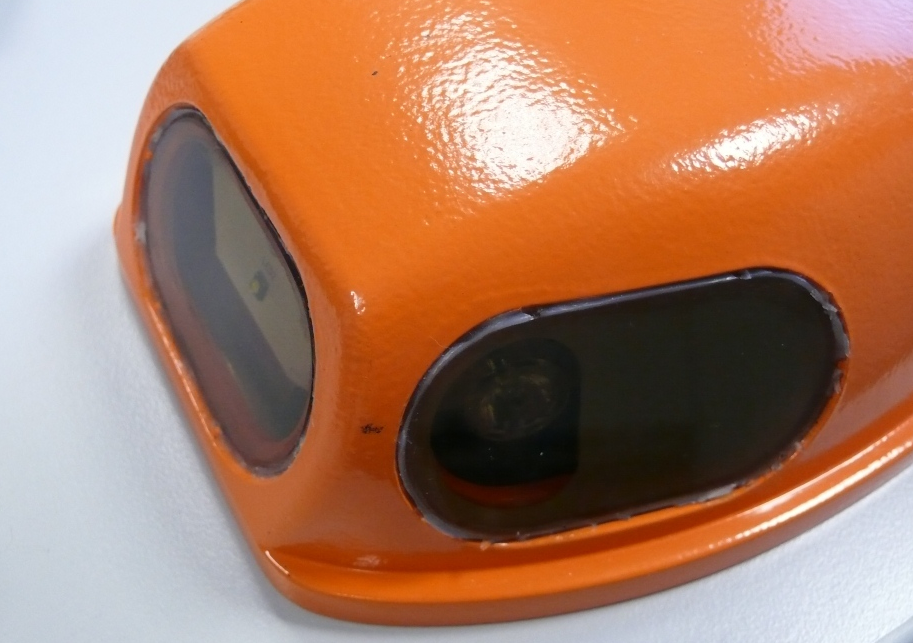}\label{Fig:vendor1glass4kGy}}
	\hfil
	\subfloat[]{\includegraphics[height=5cm]{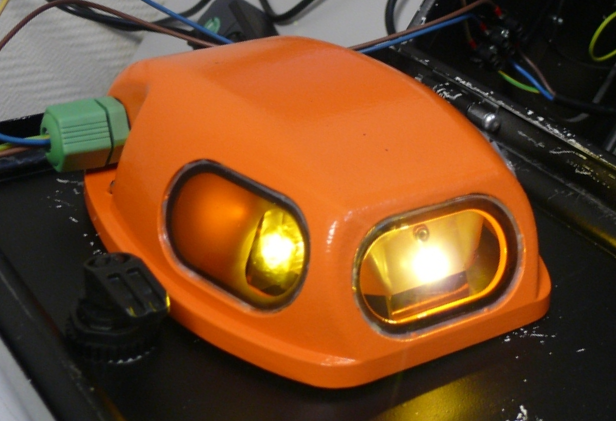}\label{Fig:vendor1glasson}}
	\caption{(a) Detail of Vendor 1 luminaire glasses after 4.1~kGy irradiation. (b) Vendor 1 luminaire on, after 50~kGy irradiation.}
\end{figure}

\subsection{Analysis of results}
\begin{figure}
	\centering
	\subfloat[]{\includegraphics[width=0.48\columnwidth]{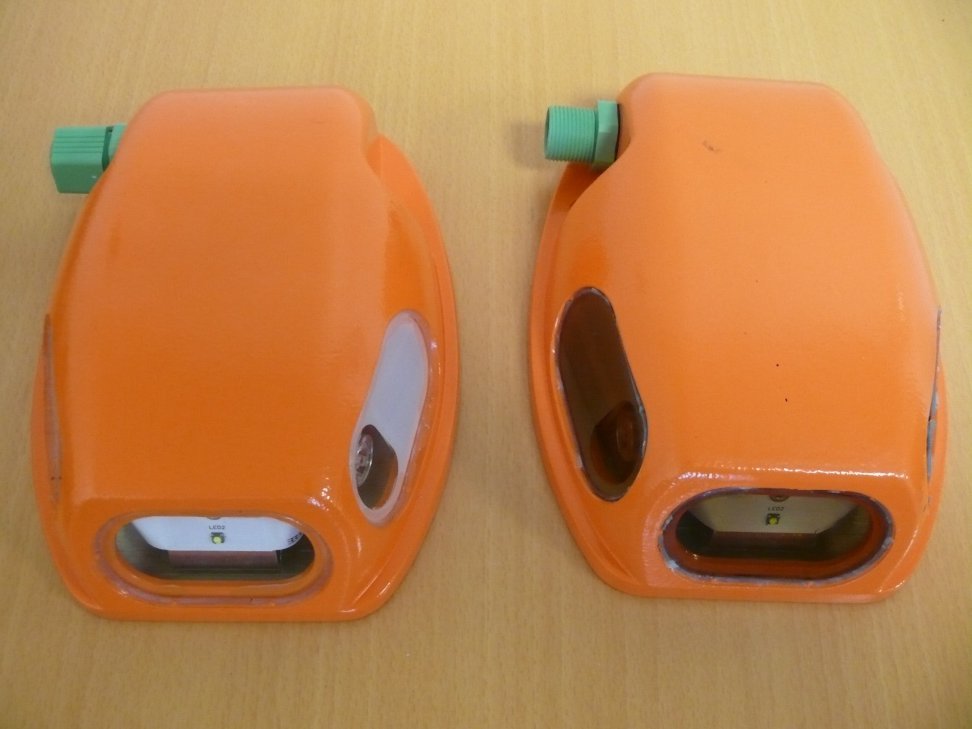}\label{fig:vend1_step1}}
	\hfil
	\subfloat[]{\includegraphics[width=0.48\columnwidth]{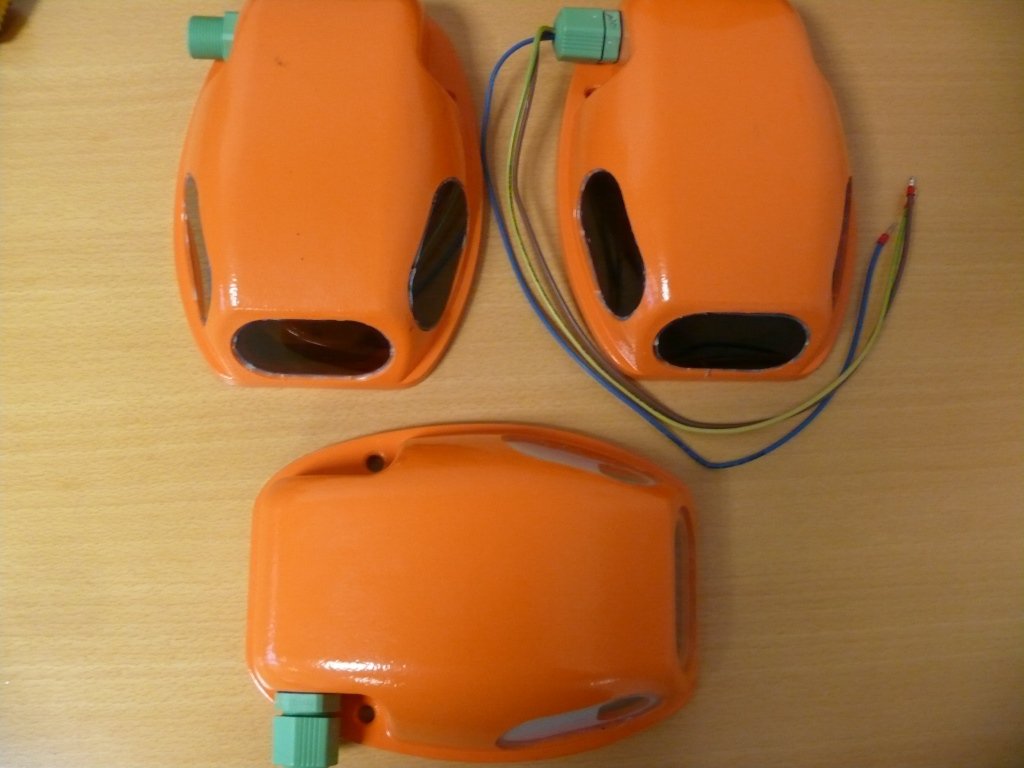}\label{fig:vend1_step2}}
	\hfil
	\subfloat[]{\includegraphics[width=0.48\columnwidth]{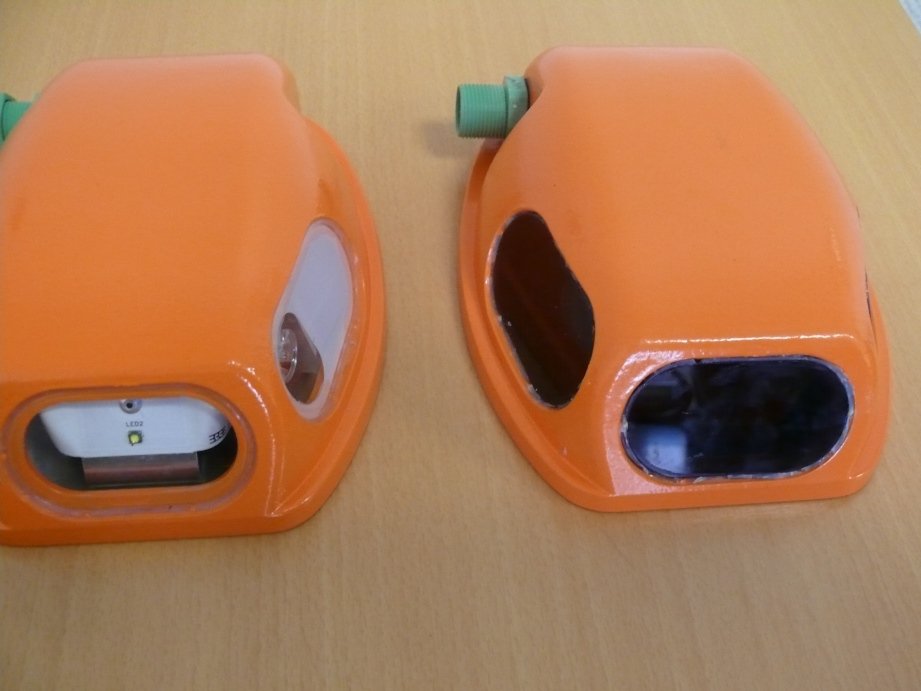}\label{fig:vend1_step3}}
	\hfil
	\subfloat[]{\includegraphics[width=0.48\columnwidth]{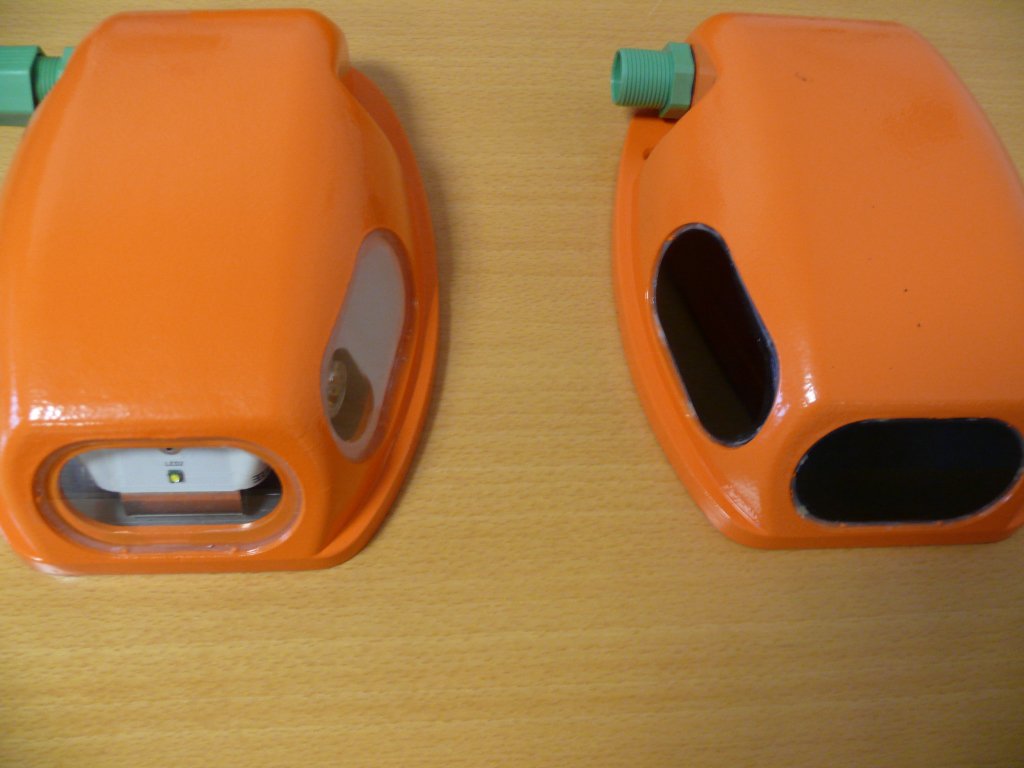}\label{fig:vend1_step4}}
	\hfil
	\subfloat[]{\includegraphics[width=0.48\columnwidth]{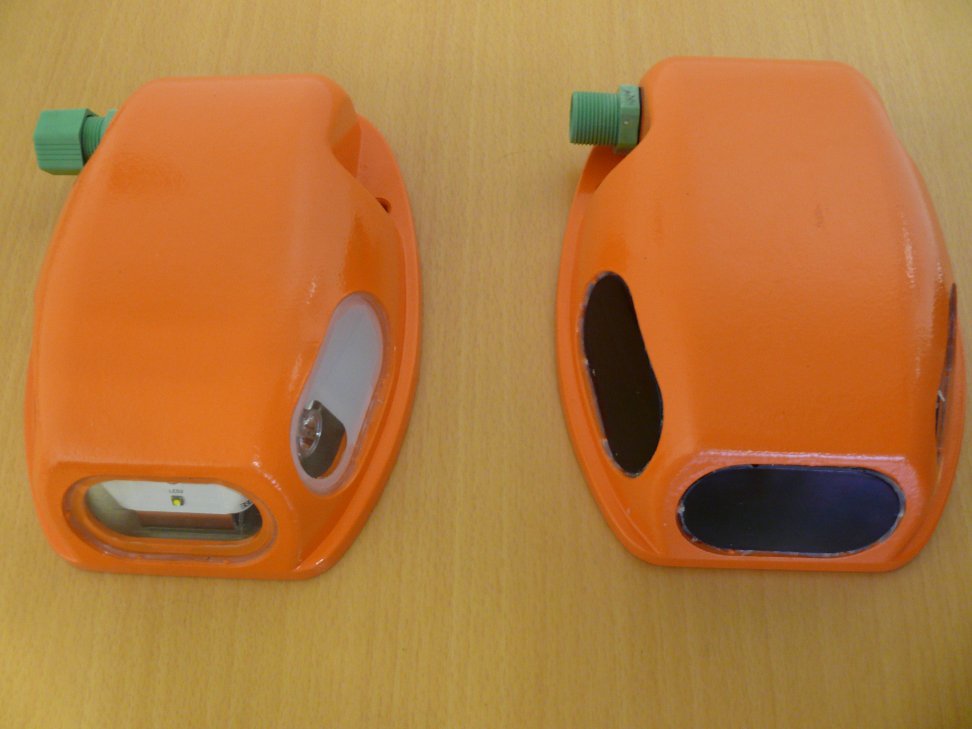}\label{fig:vend1_step5}}
	\caption{Comparison between irradiated and non-irradiated Vendor 1 luminaires after exposure to: (a) 4.1~kGy, (b) 15.3~kGy, (c) 34.2~kGy, (d) 50~kGy, (e) 100~kGy. In Figure (b) both irradiated luminaires are shown.}
\end{figure}

\begin{figure}
	\centering
	\subfloat[]{\includegraphics[width=0.48\columnwidth, angle=180, origin=c]{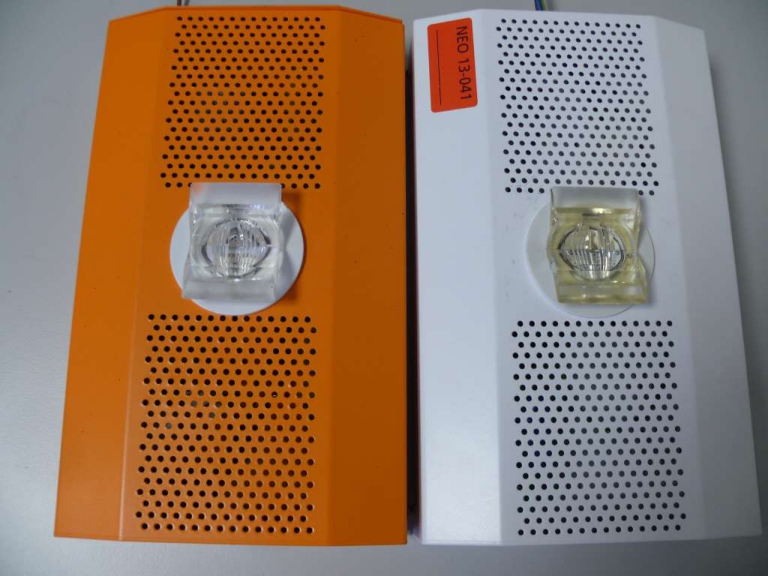}\label{fig:vend2step1}}
	\hfil
	\subfloat[]{\includegraphics[width=0.48\columnwidth]{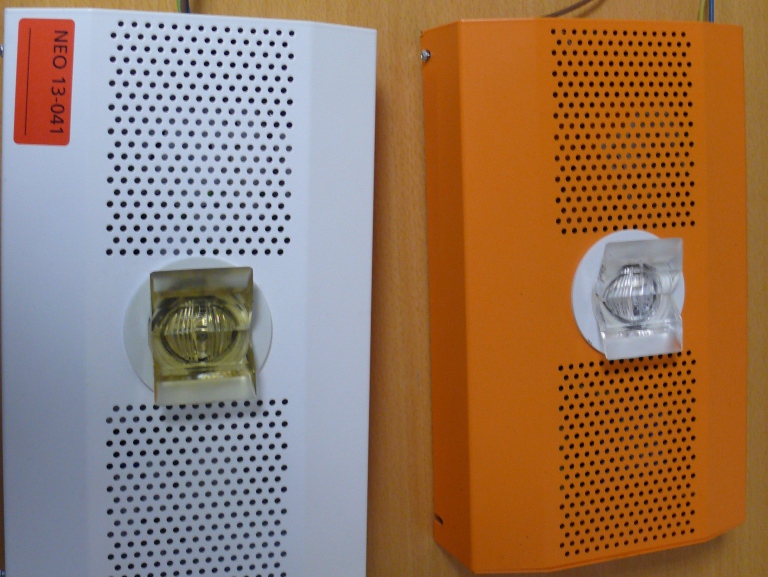}\label{fig:vend2step2}}
	\hfil
	\subfloat[]{\includegraphics[width=0.48\columnwidth]{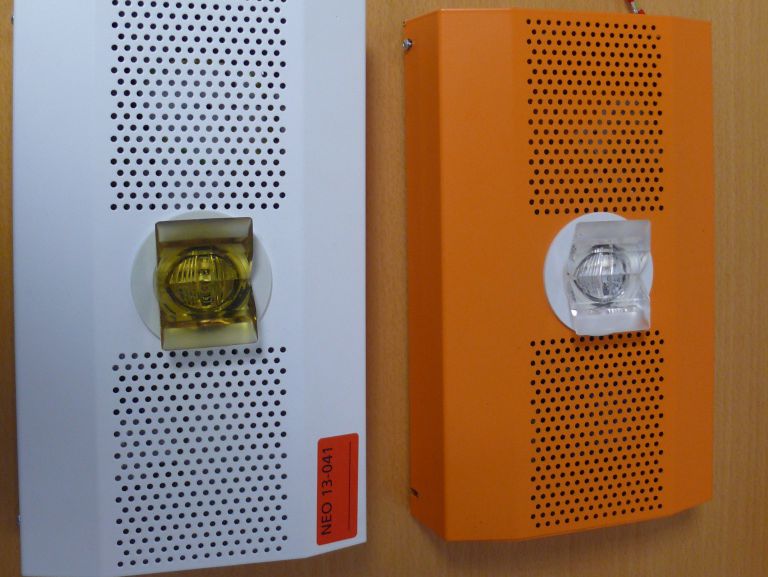}\label{fig:vend2step3}}
	\hfil
	\subfloat[]{\includegraphics[width=0.48\columnwidth]{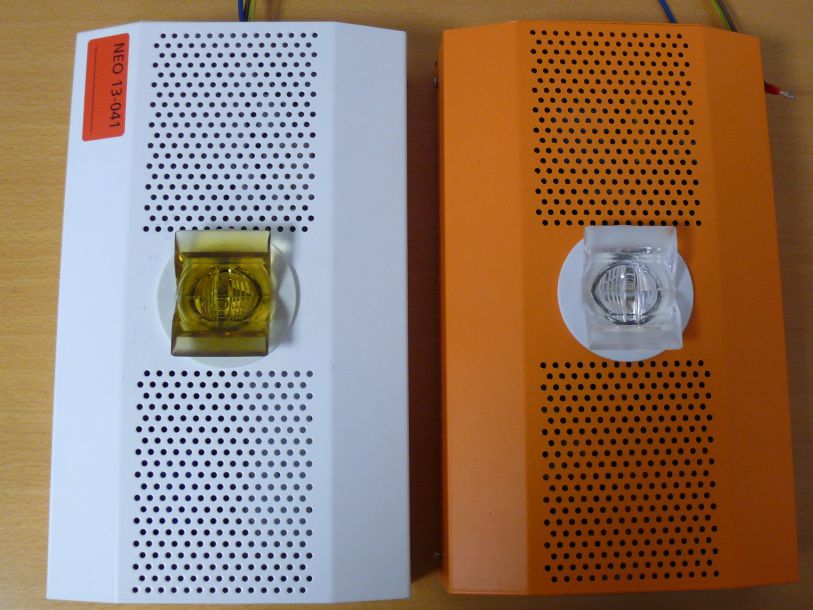}\label{fig:vend2step4}}
	\hfil
	\subfloat[]{\includegraphics[width=0.48\columnwidth, angle=180, origin=c]{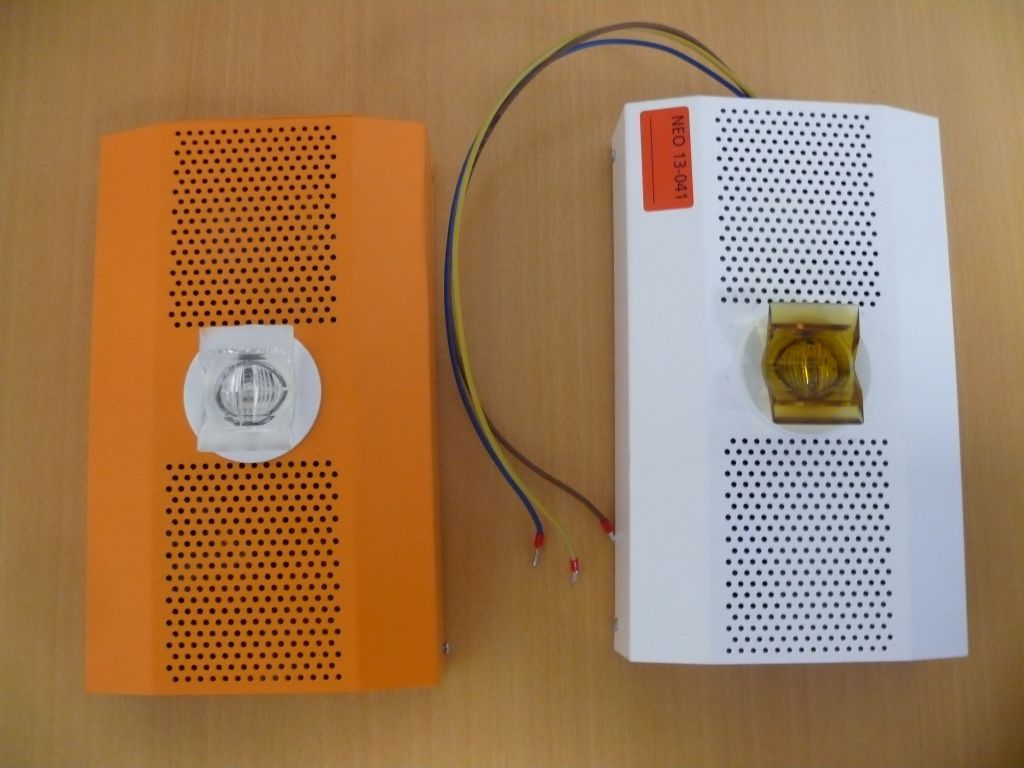}\label{fig:vend2step5}}
	\caption{Comparison between an irradiated (white) and non-irradiated (orange) Vendor 2 luminaire after exposure to: (a) 4.1~kGy, (b) 15.3~kGy, (c) 34.2~kGy, (d) 50~kGy, (e) 100~kGy.}
\end{figure}

From a comparison of the luminous flux before and after irradiation, an overall reduction of 80.3\% has been observed for Vendor 1 luminaires. This reduction appears to be predominantly due to the glass, which became dark just after the 4.1~kGy exposure (see Fig. \ref{Fig:vendor1glass4kGy}), and the PMMA lenses of the LEDs, which turned yellow. Figure \ref{Fig:vendor1glasson} shows the luminaire switched on, emitting a yellowish light, after a dose of 50~kGy. Finally, Figures \ref{fig:vend1_step1}-\ref{fig:vend1_step5} show Vendor 1 luminaires after every irradiation step. 	
As for Vendor 2 product, the drop in the luminous flux is slightly lower than Vendor 1, thanks to the absence of the glass, and can be ascribed to the degradation in the PMMA which started to become yellow after a dose of 15~kGy. Figures \ref{fig:vend2step1}-\ref{fig:vend2step5} illustrate Vendor 2 luminaires after every irradiation step.
The variations in luminous efficacy are close to those in luminous flux for both products, meaning that no significant variation in input power (due to LEDs and diodes degradation) has been observed.

\begin{figure}[!t]
	\centering
	\subfloat[]{\includegraphics[width=0.65\columnwidth]{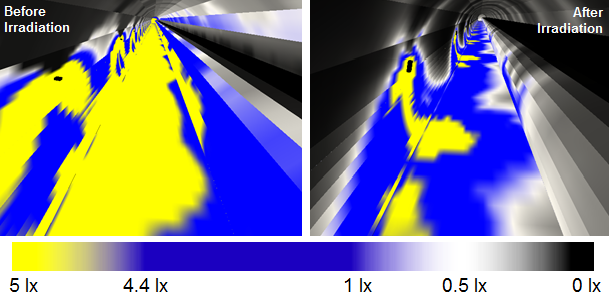}\label{Fig:visualafter}}
	\hfil
	\subfloat[]{\includegraphics[width=0.65\columnwidth]{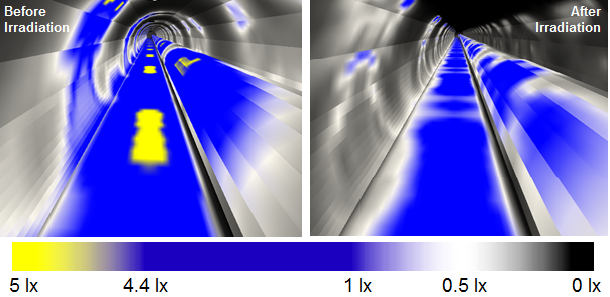}\label{Fig:visualafter_vend2}}
	\hfil
	\subfloat[]{\includegraphics[width=0.65\columnwidth]{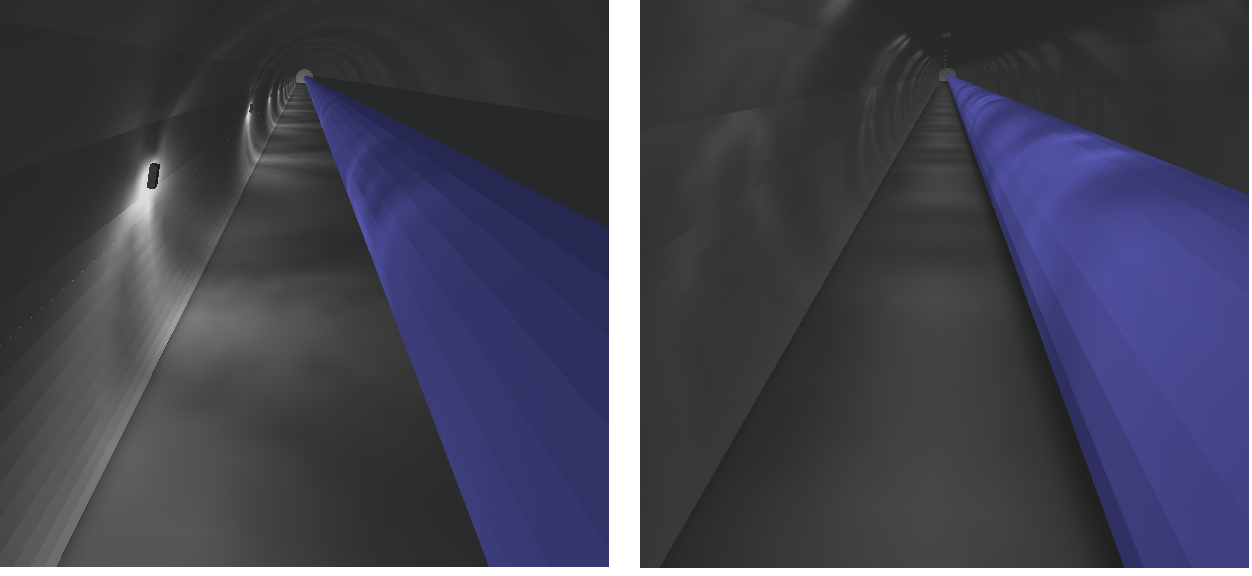}\label{Fig:visualnormal}}
	\caption{(a) 3D rendering of the DIALux simulation with Vendor 1 luminaires, before (left) and after (right) 100~kGy exposure. (b) 3D rendering of the DIALux simulation with Vendor 2 luminaires, before (left) and after (right) 100~kGy exposure. (c) Natural visual impression using Vendor 1 (left) and Vendor 2 (right) luminaires after 100~kGy exposure.}	
\end{figure}

The data from the photometric measurement has been integrated into a DIALux model for a lighting simulation of a typical segment of the LHC tunnel. Fixing centres of the luminaires are designed to account for the reduction in luminous flux due to radiation. In the simulation, Vendor 1 luminaires are placed on the tunnel wall at fixing centres of 14~m, while Vendor 2 luminaires are positioned on the ceiling of the tunnel at 10~m fixing centres. The simulation results of Vendor 1 product are summarized in Table \ref{table:dialuxVendor1and2} 
and a 3D rendering in false colour is shown in Fig. \ref{Fig:visualafter}.
We recall that EN 1838 standard requires a minimum of 0.5 lx for anti-panic lighting and 1 lx for escape routes (at centre line). For this reason, the plastic and glass types for future luminaires will be carefully selected to improve optical transmission when irradiated. The same analysis has been performed for Vendor 2 luminaire, and results are shown in Table \ref{table:dialuxVendor1and2} and Fig. \ref{Fig:visualafter_vend2}. Figure \ref{Fig:visualnormal} illustrates the natural visual impression of the tunnel segment using luminaires of both vendors after 100~kGy irradiation.

\begin{table}[!h]
	\renewcommand{\arraystretch}{1.3}
	\caption{DIALux simulations of Vendor 1 and Vendor 2 luminaires in LHC tunnel.}
	\label{table:dialuxVendor1and2}
	\centering
	\begin{tabular}{lcccccc}
		\toprule
		\multirow{2}{*}{Quantity} & \qquad & \multicolumn{2}{c}{Vendor 1}  & \qquad &\multicolumn{2}{c}{Vendor 2} \\ 
			& \qquad & Before & After  & \qquad &  Before & After \\ \midrule
		Average over the whole surface & \qquad & 9.78~lx & 2.01~lx & \qquad & 3.46~lx & 1.60~lx \\ 
		Maximum value  over the whole surface & \qquad & 17.0~lx & 9.45~lx & \qquad& 6.01~lx & 2.78~lx\\
		Minimum value along the escape route & \qquad & 5.67~lx & 0.65~lx & \qquad & 1.85~lx & 0.92~lx \\ 
		Minimum value at center line along escape route & \qquad & 7.48~lx & 1.22~lx & \qquad &  2.02~lx & 1.02~lx  \\ \bottomrule
	\end{tabular}
\end{table}

\section{Future developments}
The need for a new PSU as the first stage in radiation hardening COTS luminaires, together with impending obsolescence of the GBU8K rectifier, has spurred the production of a reference design, released under the CERN Open Hardware License \cite{devine2016psu}. Future tests up to 100~kGy are planned for the new PSU in CHARM facility at CERN \cite{ThorntonCHARM2016} using mixed field radiation produced from 24 GeV protons, investigating also the effect of different load conditions. Dedicated irradiation tests for the power LEDs are scheduled at IRRAD facility at CERN \cite{RavottiIRRAD2014}, so as to identify the impact of displacement damage using 24 GeV protons, under different bias conditions. Likewise, gamma ray irradiation tests for samples of optical components (PMMA lenses and borosilicate windows) are planned. Future studies will address the analysis of photometric data from the neutron irradiated sample luminaires, which will provide a stronger indication on the effect of displacement damage to the LEDs, and the consequent impact on overall system performance. For the purposes of this paper, the effects of annealing within the LED die have been ignored due to a significant delay between irradiation and photometric measurement, however the impact of this phenomena would also be an interesting subject for future study, within the context of the duty cycle for the luminaires and operational practices of particle accelerators. 

\section{Conclusions}
In this paper, we presented progress towards radiation hardening of LED emergency luminaires for use in accelerators environments. 
After the identification of the PSU as the cause of early catastrophic failures in COTS luminaires, we described a simple AC/DC power supply for LED emergency luminaires, the inclusion of which allows modified COTS luminaires to remain functional up to a dose of 100~kGy. The proposed luminaires are now being installed within limited areas of the CERN accelerator complex. Future TID tests are planned in order to bound the lifetime of the whole LED luminaire; moreover, detailed study of the principal optical components, (power LEDs, plastic lenses and glass) are scheduled. Future studies identified include active and non-active irradiation tests of the reference PSU design up to 100~kGy.

\section*{Acknowledgments}
The authors would like to thank Jean-Marie Foray for commencing the LED luminaire testing programme during his time at CERN from 2009-2014.

\bibliographystyle{ieeetr}
\bibliography{RadHardLighting}

\end{document}